\documentclass[conference]{IEEEtran}

\usepackage{xspace}
\usepackage{graphicx}
\usepackage{pifont}
\usepackage{algorithmic}
\usepackage{algorithm}
\usepackage{float}
\usepackage{moreverb}
\usepackage{booktabs}
\usepackage{multirow}
\usepackage{subfigure}
\usepackage{color}
\usepackage{etoolbox}
\usepackage{arydshln}
\usepackage{caption}
\usepackage{url}
\usepackage{subfig}

\graphicspath{{.}{./}}

\newcommand{\xref}[1]{Section~\ref{#1}}

\newcommand{\fref}[1]{Figure~\ref{#1}}
\newcommand{\tref}[1]{Table~\ref{#1}}

\newcommand{\first}{First,~}
\newcommand{\second}{Second,~}
\newcommand{\third}{Third,~}

\newcommand{\ie}{i.\,e., \@}
\newcommand{\eg}{e.\,g., \@}

\newcommand{\etal}{et~al.\xspace}
\newcommand{\perc}{\,\%\xspace}

\newcommand{\ispa}{\textsf{\begin{footnotesize}ISP08\end{footnotesize}}\xspace}
\newcommand{\ispb}{\textsf{\begin{footnotesize}ISP10\end{footnotesize}}\xspace}
\newcommand{\wow}{\textsf{\begin{footnotesize}World of Warcraft\end{footnotesize}}\xspace}

\newcommand{\lions}{\textsf{\begin{footnotesize}Lions\end{footnotesize}}\xspace}
\newcommand{\tigers}{\textsf{\begin{footnotesize}Tigers\end{footnotesize}}\xspace}
\newcommand{\wm}{\texttt{Wm}\xspace}
\newcommand{\wms}{\texttt{Wm/s}\xspace}

\begin{document}

\title{Tigers vs Lions: Towards Characterizing\\Solitary and Group User Behavior in MMORPG}

\author{
\IEEEauthorblockN{
Juhoon Kim\IEEEauthorrefmark{1}\IEEEauthorrefmark{2},
Nikolaos Chatzis\IEEEauthorrefmark{1}\IEEEauthorrefmark{2},
Matthias Siebke\IEEEauthorrefmark{2},
Anja Feldmann\IEEEauthorrefmark{1}\IEEEauthorrefmark{2}}
\IEEEauthorblockA{\IEEEauthorrefmark{2}Technische Universit{\"a}t Berlin, \IEEEauthorrefmark{1}Telekom Innovation Laboratories}
\IEEEauthorblockA{\{jkim,nikolaos,anja\}@net.t-labs.tu-berlin.de\IEEEauthorrefmark{1}\IEEEauthorrefmark{2}, siebke.matthias@gmail.com\IEEEauthorrefmark{2}}}

\maketitle

\begin{abstract}
The development of Internet technologies enables software developers to
build virtual worlds such as Massively Multi-Player Online Role-Playing
Games (MMORPGs). The population of such games shows super-linear growing
tendency. It is estimated that the number of Internet users subscribed
in MMORPGs is more than 22 million worldwide~\cite{url-mmodata}. However,
only little is known about the characteristics of traffic generated by
such games as well as the behavior of their subscribers. In this paper,
we characterize the traffic behavior of \wow, the most subscribed
MMORPG in the world, by analyzing Internet traffic data sets collected
from a European tier-1 ISP in two different time periods. We find that
\wow is an influential application regarding the time spent by users
(1.76 and 4.17 Hours/day on average in our measurement), while its traffic
share is comparatively low ($<$ 1\perc). In this respect, we look at the
\wow subscriber's gaming behavior by categorizing them into two different
types of users (solitary users and group users) and compare these two
groups in relation to the playing behavior (duration as the metric) and
the in-game behavior (distance as the metric).
\end{abstract}

\section{Introduction}
\label{sec:introduction}
Ever since the personal computer was introduced game developers have been 
building virtual worlds. While the hardware resources,~\eg{CPU/GPU power
and memory capacity}, have been the key constraints on developing such games
in the past, network resources and properties,~\eg{bandwidth and latency}, play
an important role in today's online games. Given the fact that the number of
Massively Multi-player Online Role Playing Game (MMORPG) subscribers rapidly
increases every year and that it reached 22 million in 2011~\cite{url-mmodata},
it is crucial for ISPs to understand characteristics of Internet traffic
generated by MMORPGs and the playing behavior of their users.

According to several reports (see for instance~\cite{url-wow,url-mmodata}),
\wow (a.k.a. WoW) alone accounts for more than 50\perc of the total MMORPG
subscriptions. Thereby, this particular online game has attracted much
attention of academia and industry. 

In this paper, we present the analysis of MMORPG traffic and users' gaming
behavior focusing on \wow as a representative game. The contributions of our
paper are mainly three-fold. \first we reveal the characteristics of \wow game
traffic by thoroughly analyzing Internet traffic collected from a European
tier-1 ISP. \second we present changes of traffic behavior by observing two
data sets collected in different time periods. \third we compare the gaming
behavior of users who play the game alone and users who play the game together
with other users behind the same middle box. As far as we know, none of the
previous studies~\cite{kihl10-aow,john09-ami,pittman07-ams,szabo09-oti,matteo08-itl}
questioned on the theme (the user behavior of MMORPGs) from this perspective.

The remainder of this paper is structured as follows. \xref{sec:background}
provides the necessary background for understanding \wow. We discuss related
work in~\xref{sec:relatedwork}. The implementation of our protocol analyzer is
discussed in~\xref{sec:methodology} and traffic traces used for this study are
described in~\xref{sec:datasets}. \xref{sec:trafficbehavior} and
\xref{sec:userbehavior} explore characteristics of \wow traffic and \wow
subscriber's gaming patterns, respectively. Finally, we conclude the paper
in~\xref{sec:conclusion}.

\section{Overview of World of Warcraft}
\label{sec:background}
\wow was first introduced in the market by Blizzard Entertainment in 2004 and
soon became one of the most subscribed (the number of its worldwide
subscriptions reached 12 million in October 2010~\cite{url-wow}) online games.
A user\footnote{The term "user", "subscriber", and "player" are interchangeably
used in this paper.} within \wow is represented as a graphical form, which is
often called the avatar, whose identity (name, gender etc.) is usually different
from the user's real world identity. This type of game is commonly referred to
as a Massively Multi-player Online Role Playing Game (MMORPG).

Unlike traditional computer games, in which the user plays as the only intellectual
game entity and the other entities act based on the story line programmed by the
game developers, MMORPGs provide their subscribers virtual worlds in which users
meet and communicate with other users. This makes interactions among users to play
more important role than the story line. For MMORPGs, more than a thousand users
play the game at the same time and they build the real-world-like society in the
game world. Moreover, users own private possessions and even trade them with other
users in virtual world currency. These sorts of social features allow MMORPG
providers to introduce a different business model from that of traditional games.
While users pay for the purchase of traditional computer games, MMORPG users pay
fee based on time they play.

Although \wow is designed to accommodate a massive number of players, it needs
a certain level of load balancing. To this end, Blizzard Entertainment provides
multiple copies of the virtual world for their users, which are called realms.
Thus, avatars are only able to interact with the other avatars which are within
the same realm.

\section{Related Work}
\label{sec:relatedwork}
Previous studies in MMORPGs have mainly focused either on characterizing
gaming traffic behavior~\cite{cevizci06-aom,chen05-gta,suznjevic09-mpa} or
on determining the users' playing patterns
\cite{kihl10-aow,john09-ami,pittman07-ams,szabo09-oti,matteo08-itl}. Although
some of our findings overlap studies of the former category, we emphasize that
our work falls more into the latter category as the novelty of our findings
mainly lies on the gaming behavior analysis of solitary users and group users.

Szab\'{o}~\etal~\cite{szabo09-oti} studies how gaming traffic is influenced by
various in-game activities of game players by deeply inspecting MMORPG traffic.
Suznjevic~\etal~\cite{suznjevic09-mpa} evaluates types of actions generated
by players within the virtual world. Their work aims at determining activities
of players within the game world, while our work examines properties such as
the movement of avatars within the virtual world and the playing duration of
users. Cevizci~\etal~\cite{cevizci06-aom} studies self-similarity of online
game traffic, but they do not particularly focus on MMORPGs.
Kihl~\etal~\cite{kihl10-aow} reports that 20\perc of the households in their
measurement environment (a Swedish broadband access network with 12K users)
has active \wow players and that their average playing duration is 2.3 hours
per day. Our work is complementary to their work since we perform our
measurement based on the traffic traces collected from the topologically
similar network in 2008 and 2010, while they conduct the measurement on
traffic collected in 2009. 

Chen~\etal~\cite{chen05-gta} analyzes ShenZhou online game traffic collected at
the server side and reports that MMORPG traffic shows irregularities due to the
players' drastically diverse gaming behaviors. However, our results suggest
that there is a high level of similarity when grouping players by the
number of co-players behind the same middle box (\ie{residential gateways}).
Varvello~\etal~\cite{matteo08-itl} studies avatars' social behavior in
Second Life. They find that approximately 0.3\perc of the total subscribers are
playing the game concurrently at any point of time and that they do not move
90\perc of the connected time. They also find that avatars in Second Life tend
to organize small groups (2 to 10 avatars). Some of these numbers are noticeably
different in our study. This is likely due to the different gaming nature between
Second Life and \wow. Pittman~\etal~\cite{pittman07-ams} studies the population
dynamics of virtual worlds over time and the players' movement patterns in the
virtual world. Miller~\etal~\cite{john09-ami} analyzes the avatars' movements
within the virtual world and they find that 5\perc of visited territory accounts
for 30\perc of all time spent. We also consider the movement as an important
metric for characterizing the player's in-game behavior, but we focus rather on
the distance avatars travel than on the pattern of the movement.
Benevenuto~\etal~\cite{benevenuto09-cub} studies the user's interaction with
other online users within various social networks (\ie{Orkut, MySpace, Hi5,
LinkedIn}). Although their study is based on online social networks, the key
question asked in their work and the one in our work are similar to each other.

\section{Methodology}
\label{sec:methodology}
In this section, we describe the implementation of the analysis tool and the
classification method used to perform our measurement. Note that we are willing
to offer our anlaysis tool to other researchers for further studies on this
subject (see also \xref{sec:datasets}).

\subsection{Analysis Tool}
For our analysis, we implement the \wow protocol
analyzer with Bro NIDS~\cite{paxson99-bas} as the code base. The reason why we
choose Bro as the basis of our analysis tool is mainly due to three unique
features of Bro. \first{Bro is designed in such a way that the transport layer
protocol analyzers are hierarchically separated from the higher layer protocol
analyzers, so that developers do not need to deal with the complexity that transport
layer protocols have (\eg{TCP stream re-assembly}).} \second{Bro supports a
specialized protocol analyzer development framework which can be translated by the
BinPac protocol parser generator~\cite{pang06-bay}. The development process is
greatly eased due to the framework.} \third{Bro provides an indigenous protocol
identification mechanism, namely the Dynamic Protocol Detection (DPD)~\cite{dreger06-dap},
which allows the analyzer to identify the protocol in the semantic manner. More
precisely, the analyzer identifies a potential protocol in the beginning of the
connection based on various classification methods (\eg{signature or well-known
network ports}) then confirms or denies the decision depending on the connection's
further behavior. This unique feature of Bro improves the accuracy of the traffic
classification.}

\subsection{Classification Method}
In order to illustrate our classification method in detail, we first provide
some level of technical information about the \wow protocol. The \wow protocol
uses TCP as its transport protocol and a client typically opens two
connections towards different servers. One connection is established between the
client and the logon server in order to authenticate the subscriber and to update
relevant server/client information such as the list of game servers and the status
of the subscriber's avatar. The other connection is established between the client
and the game server in which actual gaming information,~\eg{coordinates of the
avatar and chat messages}, is exchanged.

Our analyzer is implemented to detect the protocol in two steps. The initial step
of the protocol identification is based on examining the protocol's unique byte
pattern (signature) of the first packet of the connection. We use
\texttt{\textasciicircum\textbackslash{x}00...WoW} and
\texttt{\textasciicircum..\textbackslash{x}ed\textbackslash{x}01} as
regular expression signatures of the logon connection and the game connection,
respectively. However, due to the short signature length, relying only on this step
yields a high false-positive rate. Thus, the next step verifies if the
connection responder replies with the expected signatures. In this step, we use
\texttt{\textasciicircum\textbackslash{x}00} (logon connection) and 
\texttt{\textasciicircum..\textbackslash{x}ec\textbackslash{x}01} (game connection)
as signatures.

Even after the protocol classification, the nature of the proprietary
software that the \wow programs have and partly encrypted protocol messages make the
deep inspection of \wow traffic extremely difficult. Thus, we use unencrypted part of
protocol messages for our analysis. As we will show in~\xref{sec:trafficbehavior}, more than
60\perc of the \wow packets are coordinate information which can be translated into
human readable text.

\section{Data Sets}
\label{sec:datasets}

\renewcommand\arraystretch{1.1}

\begin{table*}[t]
\caption{Overview of anonymized packet traces.}
\begin{center}
\tabcolsep3.5mm
\begin{tabular}{|l|c|c|c|c|c|c|c|c|}
\hline
& \multicolumn{3}{|c}{Overall Traffic} & \multicolumn{5}{|c|}{\wow}             \\
\cline{2-9}
Name & Mon. Year & Duration & Volume & packets & Volume & Version & Users & Avg. Playing\\
\hline
\ispa & Aug. 2008 & 12:00 -- 12:00 & $>$ 4TB & 0.91\perc & 0.48\perc & 2.4.3 & 3.0\perc (599) & 1.76 Hours \\
\ispb & Mar. 2010 & 02:00 -- 02:00 & $>$ 4TB & 0.83\perc & 0.72\perc & 3.x.y & 1.4\perc (280) & 4.17 Hours \\
\hline
\end{tabular}
\end{center}
\label{tab:traces}
\end{table*}

We use two anonymized 24-hour packet-level traffic data sets (\ispa and \ispb)
for our study. These two traces are collected at the same aggregation point
within a European tier-1 ISP in 2008 and in 2010. The monitor, using Endace
monitoring cards, operates at the broadband access router connecting customers
to the ISP's backbone. We count more than 20K DSL lines behind our vantage
point. All confidential information such as IP address and user identification
are anonymized by Bro's integrated encryption feature. The relevant information
of our traffic data sets and the number of identified \wow subscribers are
summarized in~\tref{tab:traces}.

In the light of the fact that \wow traffic is pure controlling traffic generated
by users (e.g., by mouse button clicking and/or by key pressing), as opposed to
the media content delivery, we believe that 0.48\perc and 0.72\perc
(see~\tref{tab:traces}) of traffic contributions are worthwhile to study. Note
that traffic generated by software updates is not included.

Although our traffic data is highly anonymized, we cannot make it publicly
available since the content of messages encrypted by the \wow application is
unknown to us and may include personal information. However, the measurement can
be conducted by using any such traces since our analysis tool is available
\footnote{The base code can be downloaded from http://www.bro-ids.org/ and our
analysis tool can be obtained by contacting the corresponding author of this
paper.} to public under the same license that Bro NIDS follows (BSD).

\section{Traffic Characteristics}
\label{sec:trafficbehavior}
In this section, we first illustrate the general characteristics of \wow
traffic. Then, we present that the majority of the \wow traffic is movement
messages in which coordinates of avatars and nearby objects (\eg{non-player
characters or game items}) are delivered.

\subsection{General Characteristics}
\begin{figure*}[t]
	\begin{minipage}[h]{0.32\linewidth}
		\centering
		\subfigure[Logon connections]{
			\includegraphics[width=\linewidth]{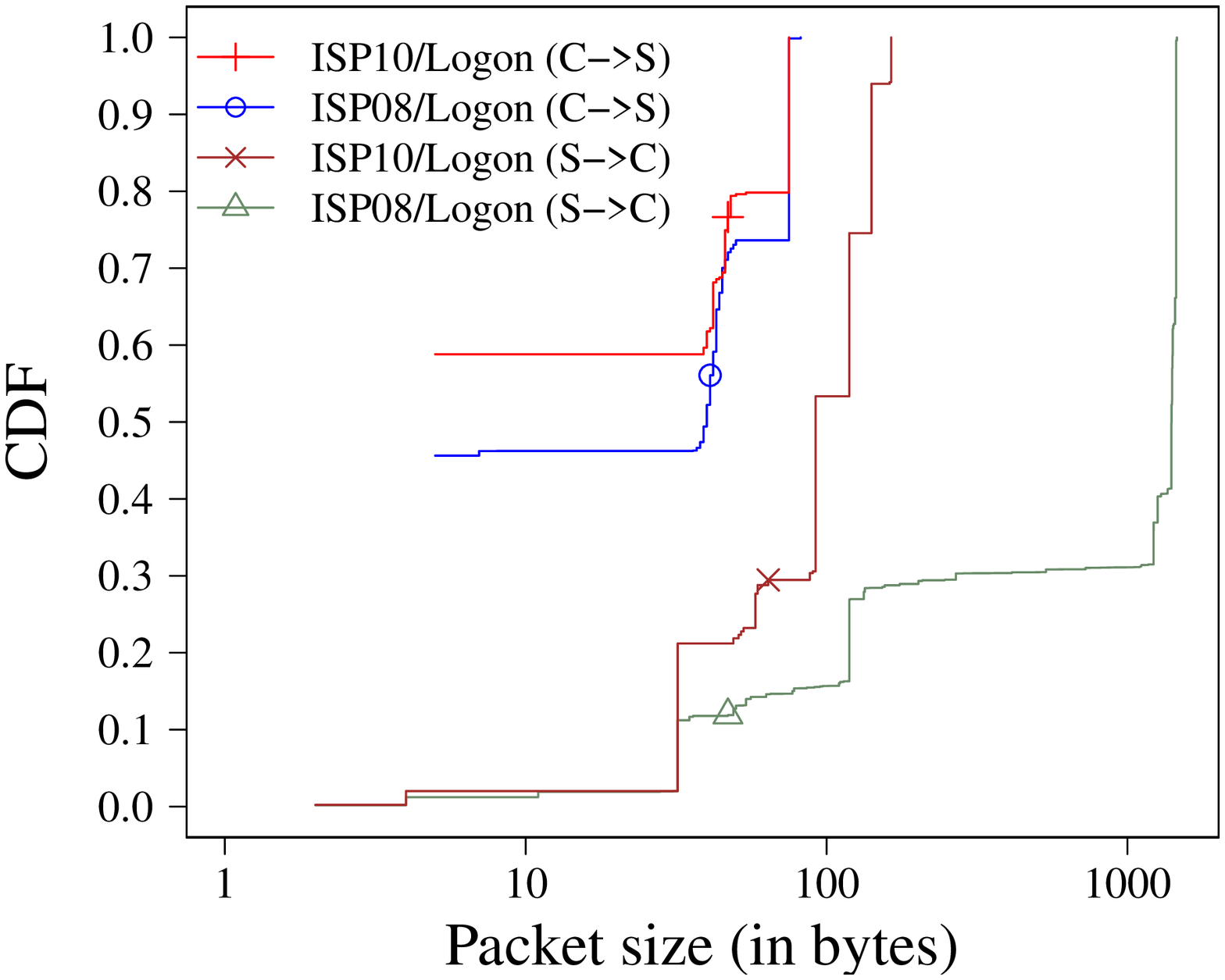}}
		\subfigure[Data connections]{
			\includegraphics[width=\linewidth]{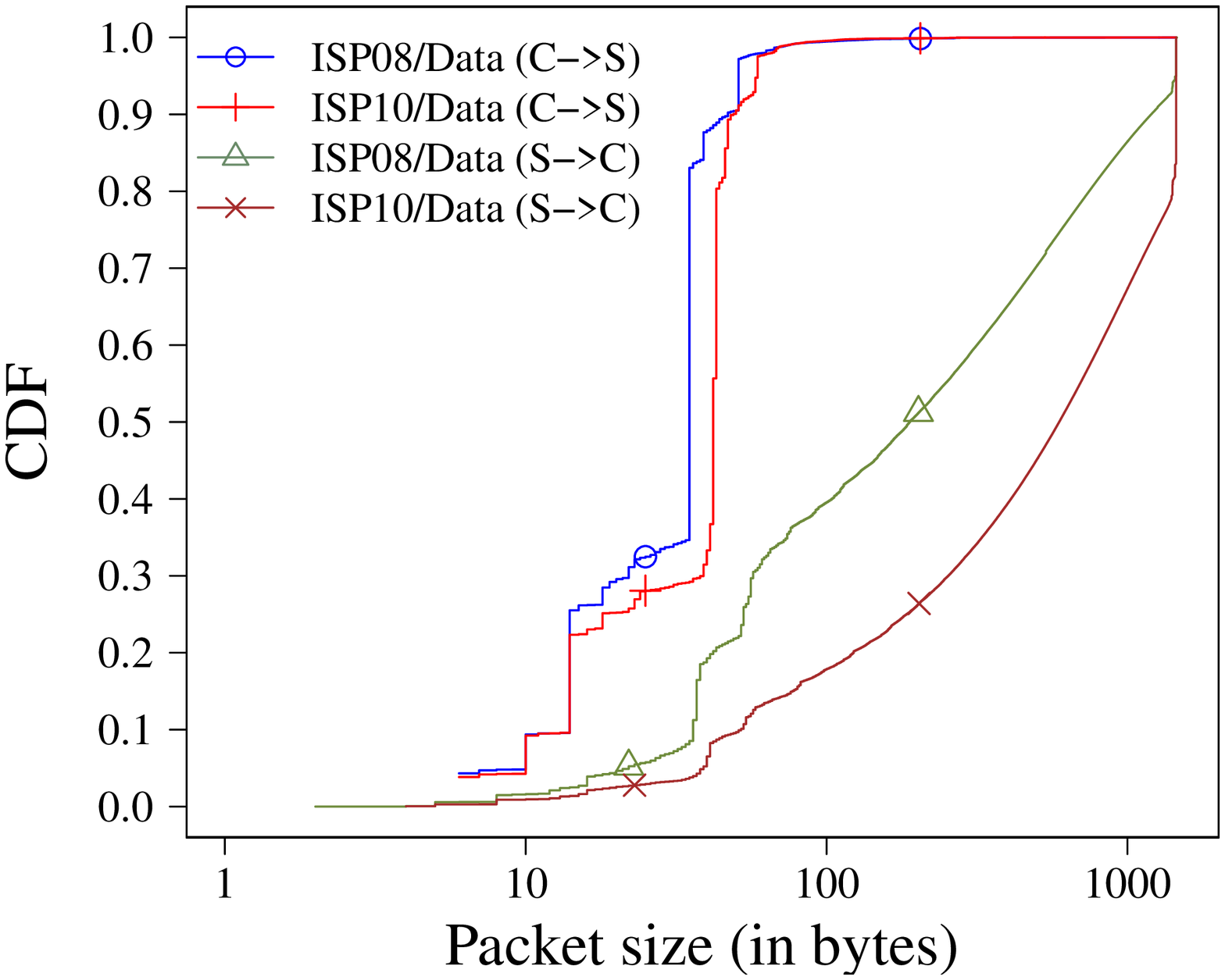}}
		\caption{Packet size}
		\label{fig:packet-size-dist}
	\end{minipage}
	\hspace{1pt}
	\begin{minipage}[h]{0.32\linewidth}
		\centering
		\subfigure[Logon connections]{
			\includegraphics[width=\linewidth]{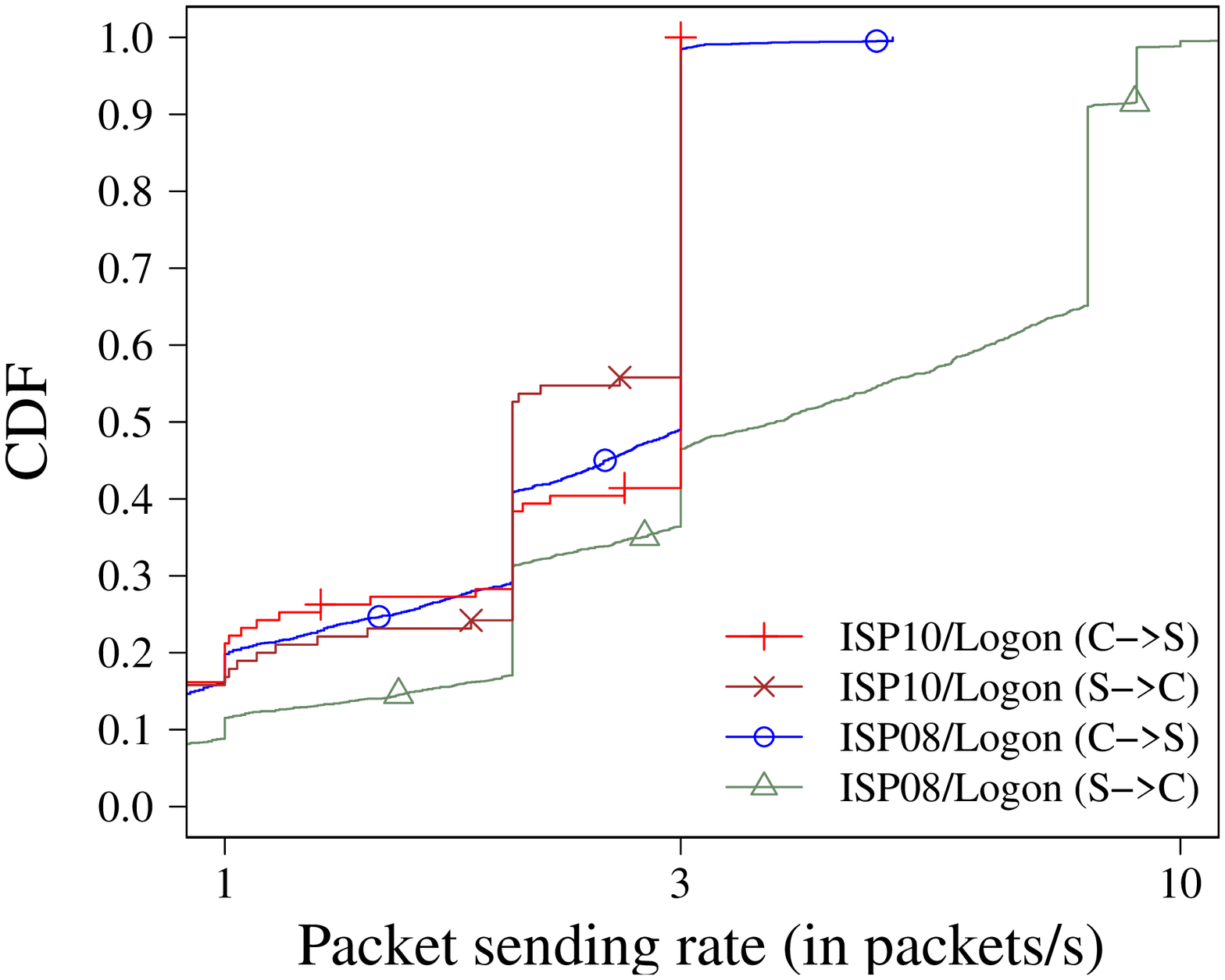}}
		\subfigure[Data connections]{
			\includegraphics[width=\linewidth]{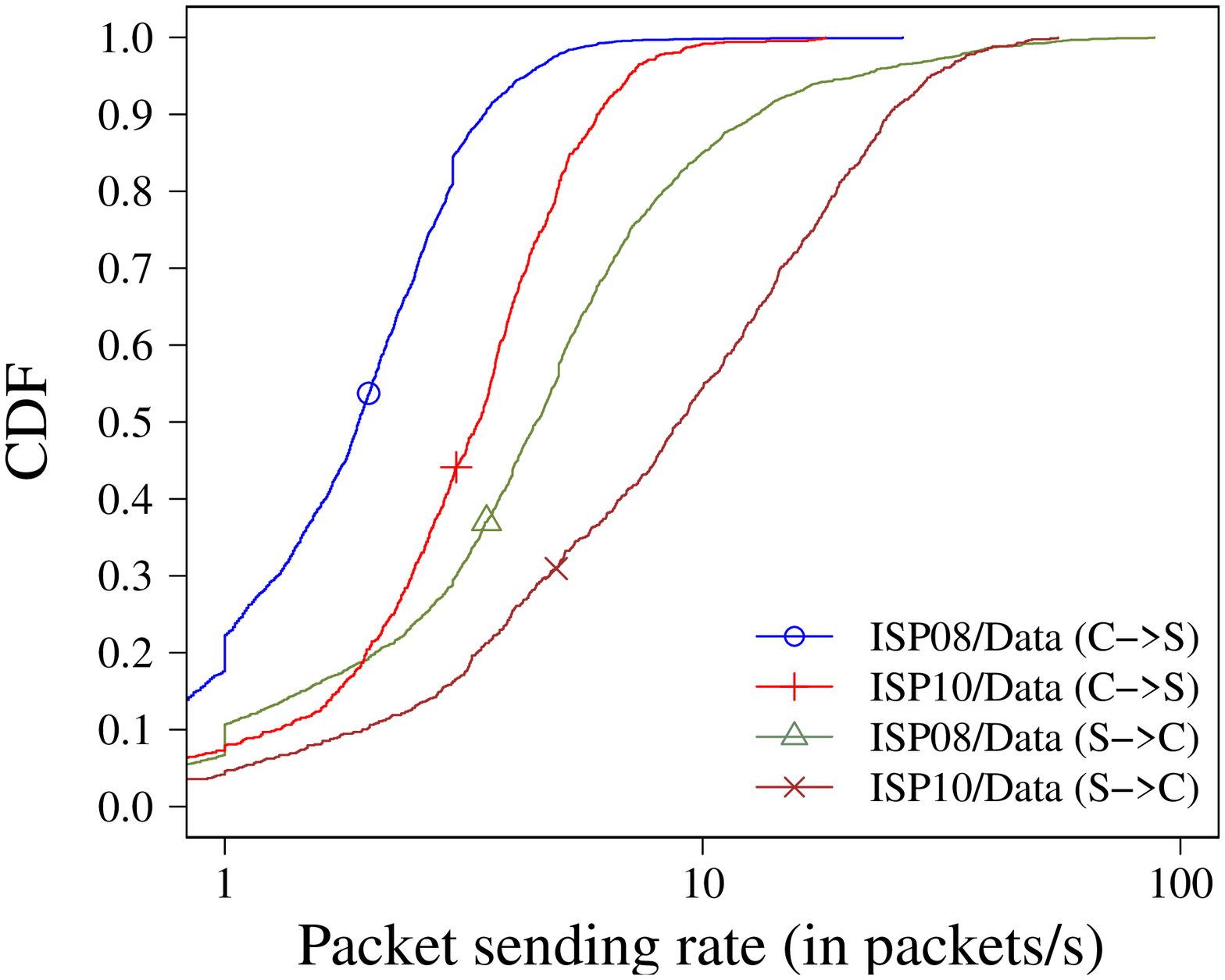}}
		\caption{Packet rate}
		\label{fig:packet-rate-dist}
	\end{minipage}
	\hspace{1pt}
	\begin{minipage}[h]{0.32\linewidth}
		\centering
		\subfigure[Logon connections]{
			\includegraphics[width=\linewidth]{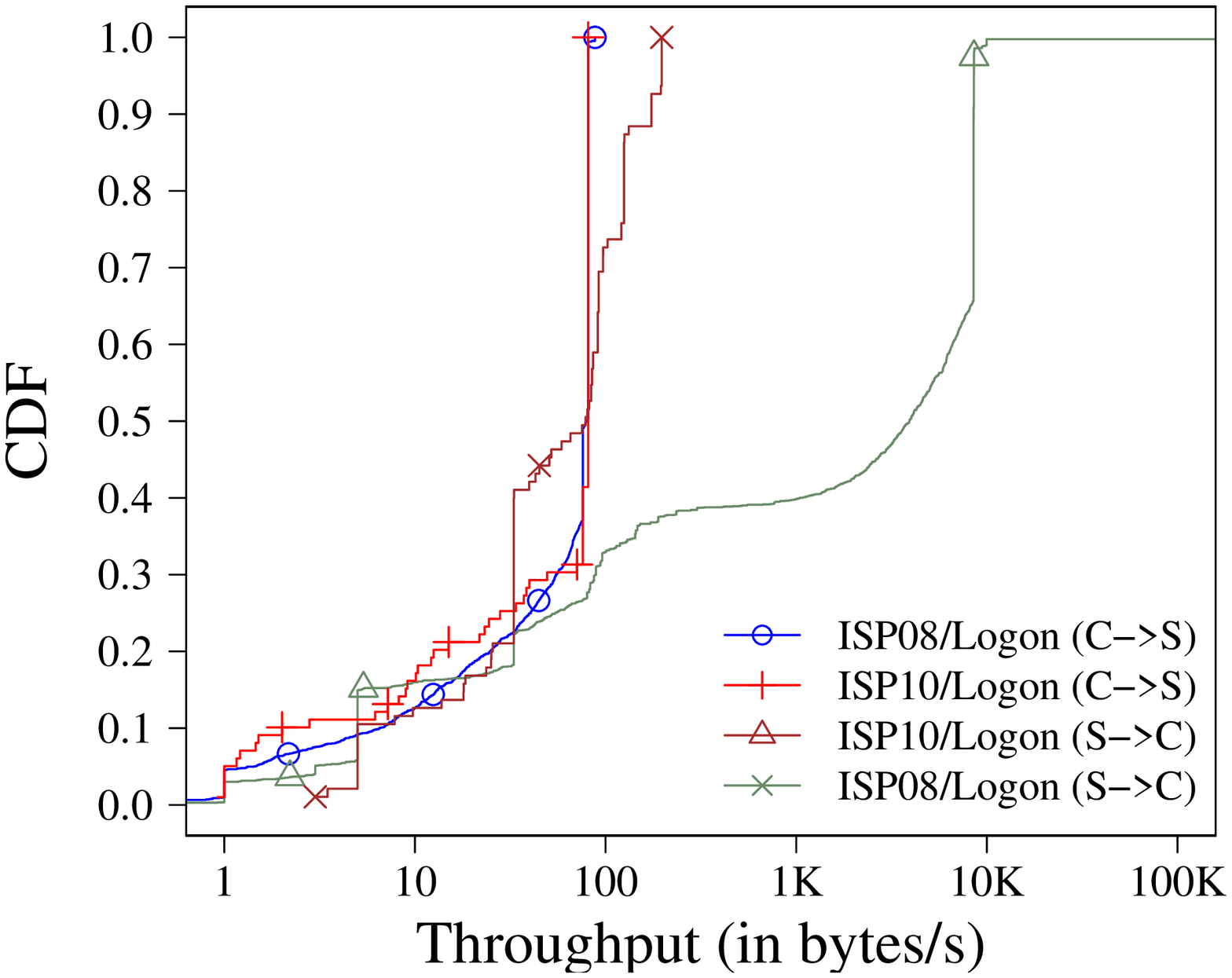}}
		\subfigure[Data connections]{
			\includegraphics[width=\linewidth]{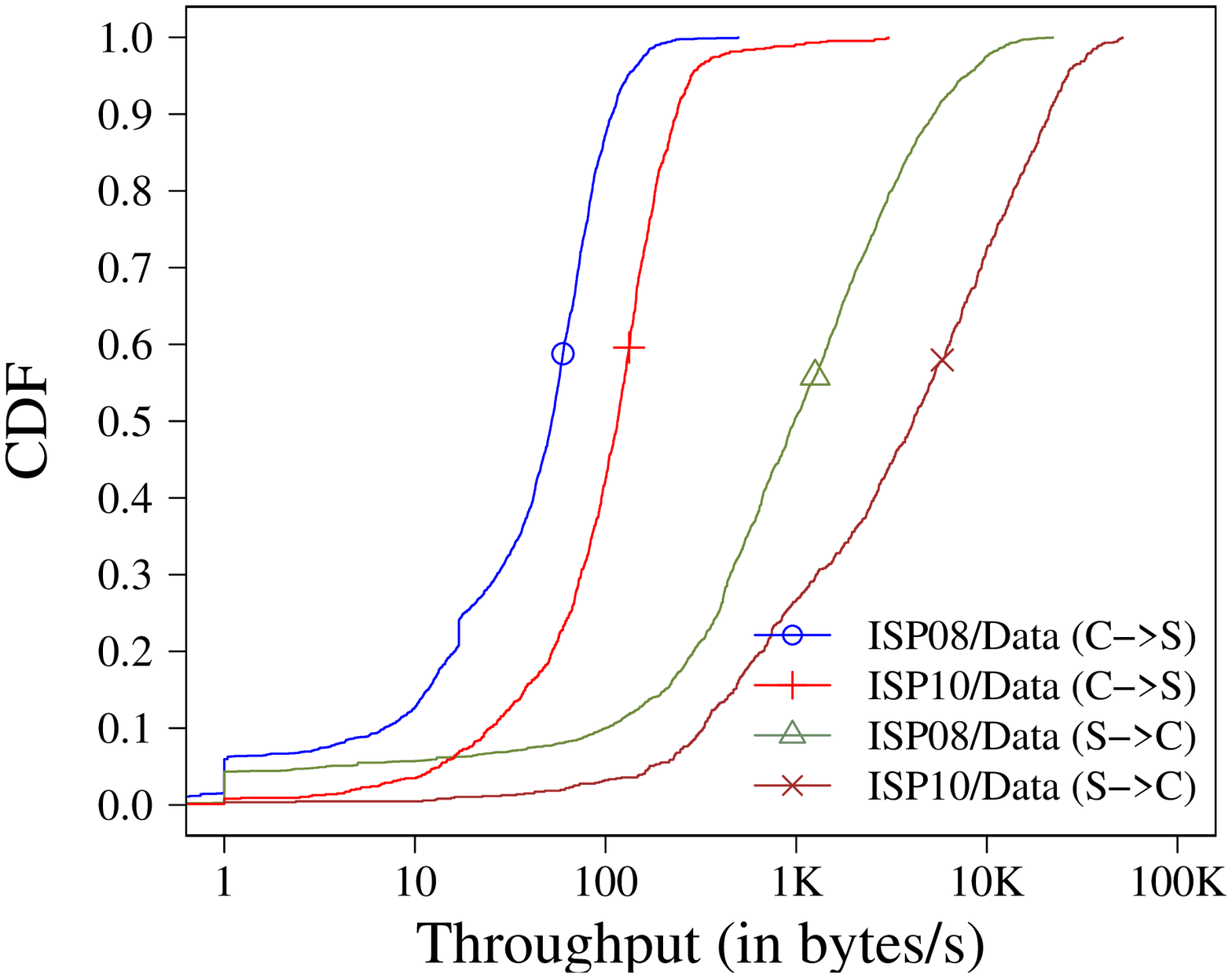}}
		\caption{Throughput}
		\label{fig:packet-throughput}
	\end{minipage}
\end{figure*}

\fref{fig:packet-size-dist},~\fref{fig:packet-rate-dist}, and
\fref{fig:packet-throughput} respectively depict the distributions of packet
size (payload only), packet sending rate, and throughput. We report results of
logon connections separately from those of data connections since those two
types of connections show clearly different behavior in terms of packet count,
traffic volume, and duration of connections. Interestingly, we observe that the
line shape of server-to-client packet streams of \ispb is significantly
different from that of \ispa in all logon plots, while the change of
client-to-server packet streams is not remarkable. Furthermore, the lines
representing data packet streams of \ispb ((b) of all three figures) have
drastically shifted compared to those of \ispa. We draw the conclusion that
the protocol has been modified during the two years in such ways that a server
delivers logon information in a more compressed or more distributed manner and
delivers gaming information more frequently.

Comparing the throughput statistics of the two traces shown in
\fref{fig:packet-throughput}, the later version of \wow protocol tends to need
more bandwidth than the earlier one. Yet, its bandwidth utilization is relatively
low considering the network bandwidth provided by today's ISPs. This means, unlike
popular belief among users, upgrading the link capacity may not be the solution
for a better gaming experience. 

\subsection{Movement Messages}
Regarding the peak of client-to-server packet streams observed in
\fref{fig:packet-size-dist} (b), 43 bytes (\ispa) and 51 bytes (\ispb) are
typical sizes identified from packets which deliver avatar's coordinates within
the virtual world to the server. The reason why the size of the movement
message differs in two traces is that the object's ID is embedded in a
different way in different versions of the protocol. This phenomenon leads us
to the conclusion that more than 60\perc of the total packets delivered from
clients to servers is the avatar's coordinate information.

By deeply inspecting server-to-client messages, we find that the majority of
messages sent from servers to clients are movement messages for updating the
client on coordinates of its nearby players and objects. However, we do not
observe the peak which is shown in client-to-server packet streams since a server
sends coordinates of various numbers of objects within one packet.

\section{Tigers vs Lions}
\label{sec:userbehavior}

\begin{table*}[thp]
\caption{Categorization of users according to the number of locally grouped co-players}
\begin{center}
\tabcolsep3.5mm
\begin{tabular}{|l|l|r|r|r|r|r|r|}
\hline
\multicolumn{2}{|c|}{Group size}  & \multicolumn{3}{c|}{\ispa}                    &     \multicolumn{3}{c|}{\ispb}               \\
\cline{3-8}
\multicolumn{2}{|c|}{(in players)} & \# of IPs.    & \# of users   & volume   &       \# of IPs   & \# of users   & volume \\
\hline
                   \tigers   &  1 & 487 (75\perc)    & 487 (54\perc)  & 53\perc    &       257 (82\perc)   & 257 (68\perc)  & 65\perc  \\
\hdashline
\multirow{4}{0.1cm}{\lions}  &  2 & 118 (18\perc)    & 236 (26\perc)  & 28\perc    &        46 (15\perc)   & 92 (24\perc)   & 30\perc \\
                             &  3 &  26  (4\perc)    &  78 (9\perc)   & 10\perc    &         8 (3\perc)    & 24 (6\perc)    & $>$ 4\perc \\
                             &  4 &  11  (2\perc)    &  44 (5\perc)   & $<$ 6\perc    &         1 ($<$1\perc) &  4 (1\perc)    & $<$ 1\perc \\
                             &  $>$ 4 $^*$ &   7  ($<$2\perc) &  55 (6\perc)   & $<$ 4\perc    &         0 (0\perc)    &  0 (0\perc)    & 0\perc \\
\hdashline
                             &  total            & 649  (100\perc)  & 900 (100\perc) & 100\perc   &     312 (100\perc)  & 377 (100\perc) & 100\perc\\
\hline
\multicolumn{8}{|l|}{$^*$ Maximum number of users behind a single IP address is 14} \\
\hline
\end{tabular}
\end{center}
\label{tab:nated}
\end{table*}

\begin{figure*}[t]
	\begin{minipage}[h]{0.32\linewidth}
		\subfigure[Connection duration]{\includegraphics[width=\linewidth]{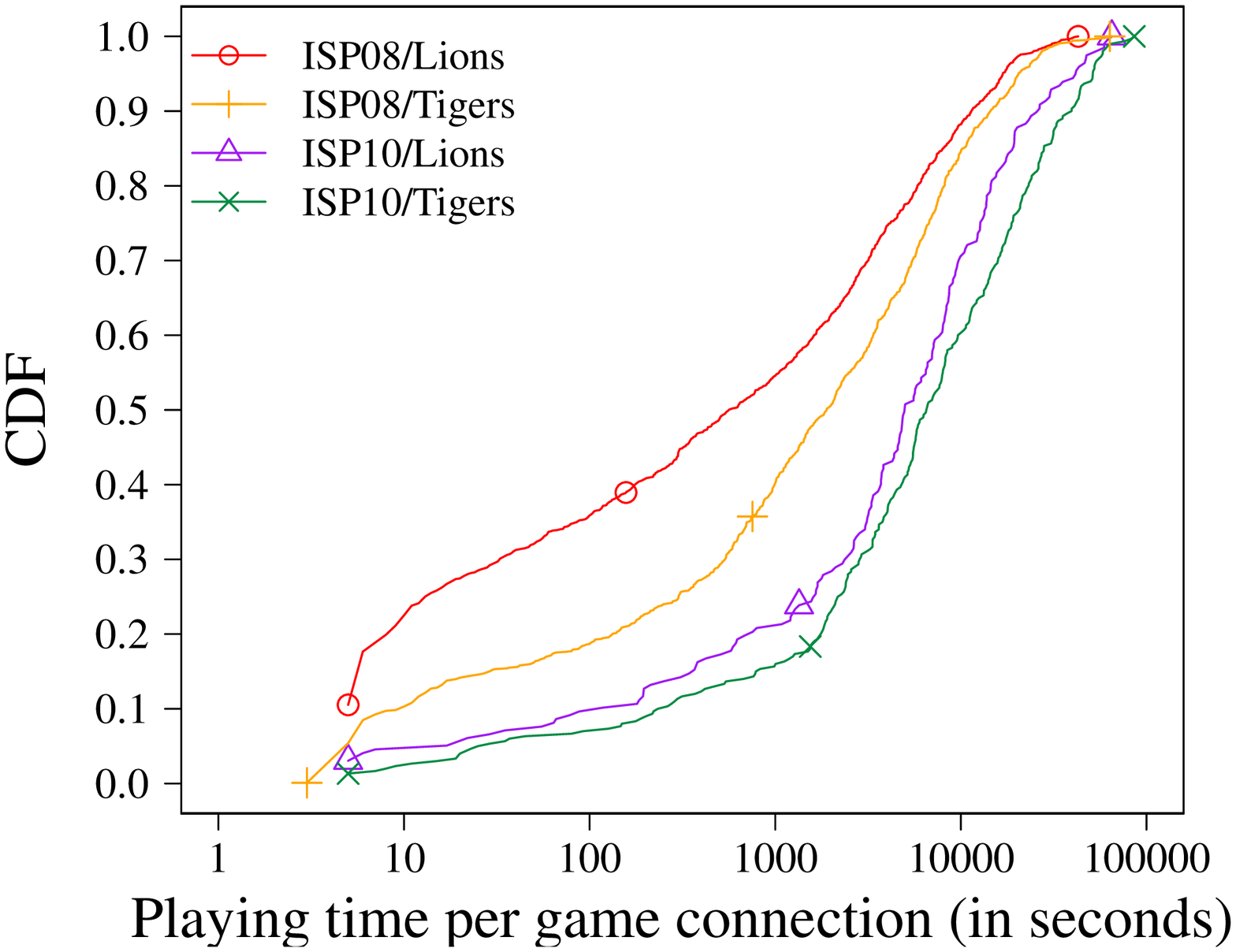}}
	\end{minipage}
	\hspace{1pt}
	\begin{minipage}[h]{0.32\linewidth}
		\subfigure[User playing time]{\includegraphics[width=\linewidth]{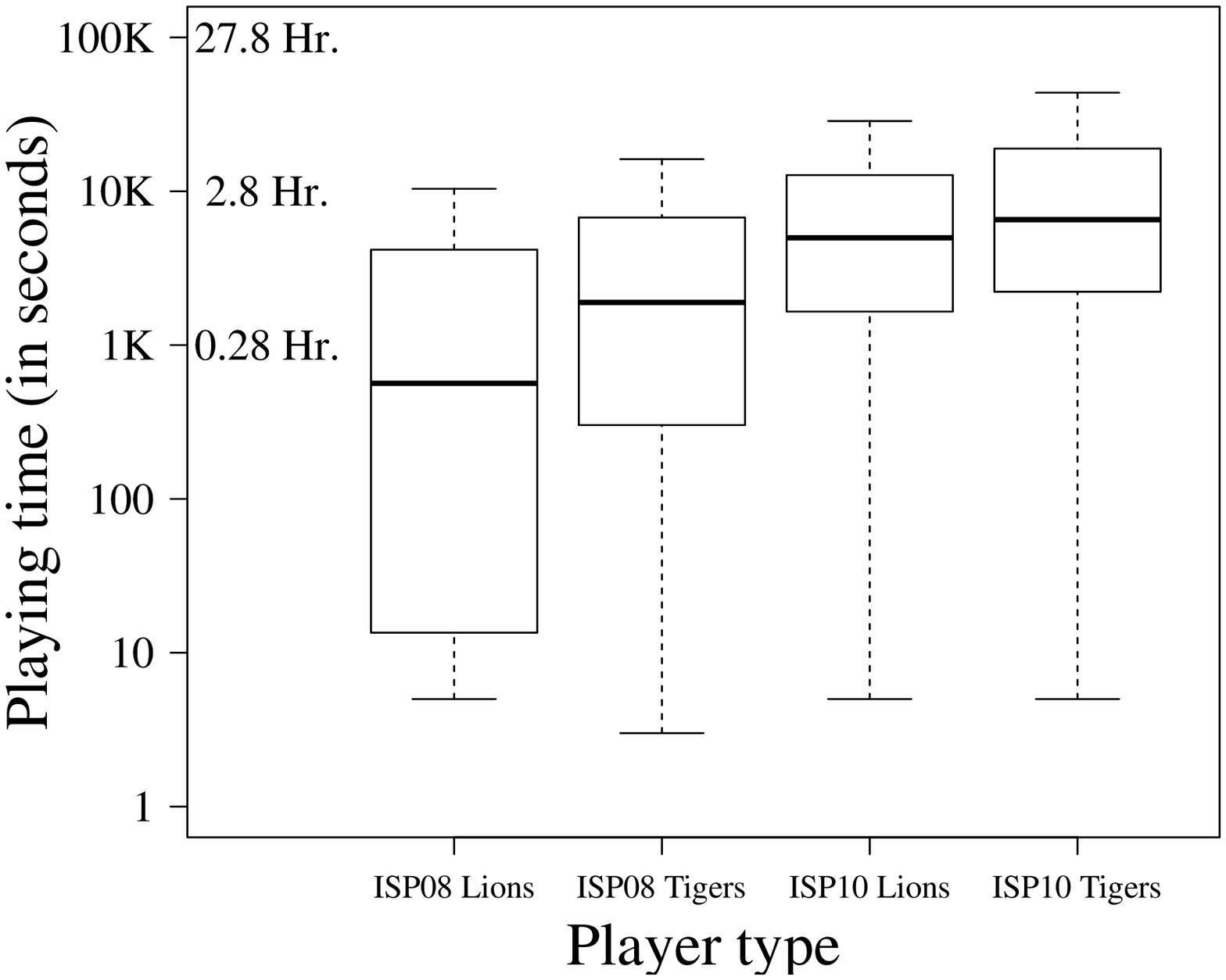}}
	\end{minipage}
	\hspace{1pt}
	\begin{minipage}[h]{0.32\linewidth}
		\subfigure[Playing time of day]{\includegraphics[width=\linewidth]{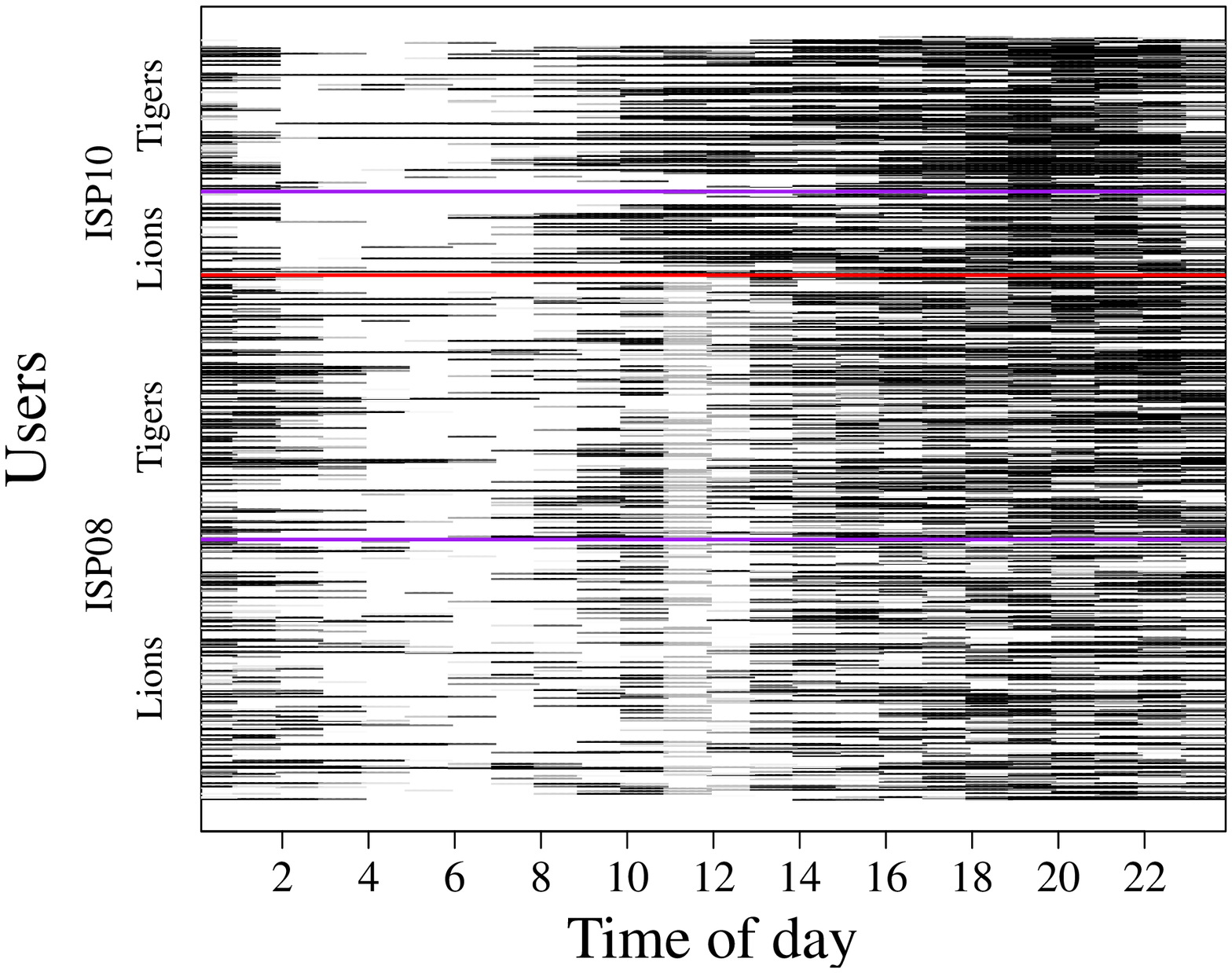}}
	\end{minipage}

	\caption{Playing time distributions}
	\label{fig:playtime}
\end{figure*}

MMORPG subscriber's gaming behavior is an often addressed research topic in
online game traffic measurement. Such studies are especially important for
ISPs in order to understand user's increasing demand for a good gaming
experience. Assuming MMORPG users as a homogeneous group, previous studies
\cite{chen05-gta,kihl10-aow,john09-ami,suznjevic09-mpa,szabo09-oti} focus
either on characterizing the overall gaming behavior of users or on comparing
the user behavior in different virtual worlds. However, towards gaining a more
complete insight into the gaming behavior of MMORPG subscribers we take a
different approach to those of earlier work.

\subsection{Playing Behavior}
We first classify \wow players into two groups. The first group consists of
users who is the only player behind an IP address. The other group consists of
users whose IP addresses are shared with other \wow users. We refer to the
former as \tigers and to the latter as \lions according to their hunting
behaviors (tigers hunt individually, while lions hunt in a pride). One must note
that players may be included in both groups multiple times due to the change of
their playing locations or the reallocation of IP addresses, thus the number
of users reported in~\tref{tab:nated} is greater than the one in
\tref{tab:traces}. It is crucial to mention that we do not focus on
characteristics of individual users, but intend to study general differences
between locally grouped players and solitary players in terms of playing time,
traffic volume, and distance their avatars move within the virtual world. We
summarize relevant statistics in \tref{tab:nated}.

With few exceptions in \ispa, we find that the number of users per group of
\lions is less than 4 which is a common maximum number of physical network ports
on home networking devices. The table explains that the fraction of \tigers
increases in \ispb.

\fref{fig:playtime} illustrates the subscriber's playing time in different
perspectives. While \fref{fig:playtime} (a) depicts duration of connections,
the result shown in \fref{fig:playtime} (b) is aggregated with the user. We
infer two observations from the figures.~\first while the number of identified
game users decreases during the two years, the playing duration tends to
increase. Taking a deeper look, we find that the fraction of users who play the
game shorter than 0.28 hours in a day decreases from 40\perc to 20\perc, while
the fraction of users who play the game longer than 2.8 hours in a day increases
from 20\perc to 40\perc.~\second considering that the y-axis of~\fref{fig:playtime}
(b) is in a logarithmic scale, solitary users (\tigers) are playing notably longer
than users playing together behind the same middle box (\lions).

In \fref{fig:playtime} (c), we illustrate the time of day that users play the game.
Note that the y-axis of this figure represents \wow users and x-axis time of day.
Each time slot is filled with various depth of gray color depending on the minutes
played in the time slot. Thus, a time slot is fully used when it is black and it is
not used when it is white. The x-axis is wrapped around the beginning time of the
measurements. It is crucial to illustrate the results in this manner in order to
make the results comparable. Indeed, the two traces do not begin at the same
time of day and also do not begin at mid night (see \tref{tab:traces}). The
sudden bright cells observed between 11 a.m. and 1 p.m. in \ispb is due to the
fact that the connections established before the beginning of the measurement
cannot be recognized by the analyzer. The figure suggests that there is only a negligible difference between
\tigers and \lions when considering the time of day they play. Unsurprisingly,
popular time slots of a day are identified between 7 p.m. and 11 p.m.

\subsection{In-game Behavior}
\begin{figure}[h!]
	\includegraphics[width=\linewidth]{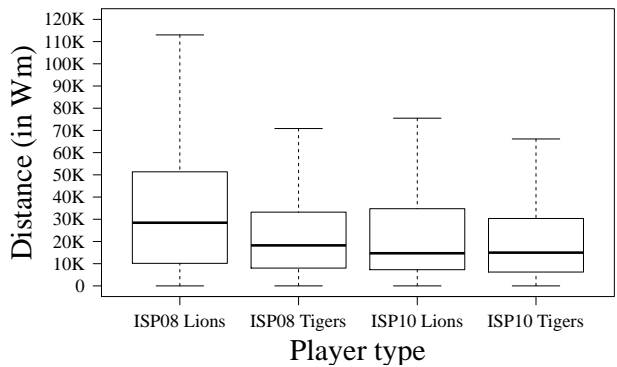}
	\caption{Movement distances of avatars}
	\label{fig:distance}
\end{figure}

\begin{figure}[h!]
	\includegraphics[width=\linewidth]{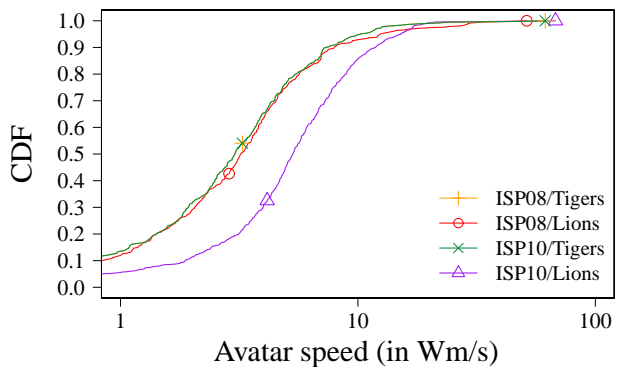}
	\caption{Speeds of avatars}
	\label{fig:speed}
\end{figure}

Next, we analyze the distance that avatars move in the virtual world during
the game play.~\fref{fig:distance} illustrates the distance distribution of the
avatar's movement. As there is no way to map the distance in the virtual
world to the real world's metric, we invent an imaginary metric \wm for measuring
the distance in \wow. In order to provide readers with the intuitional hint of this
virtual metric, we illustrate the distribution of the speed at which avatars move
in the virtual world in~\fref{fig:speed}. In this analysis, we find that about
0.8\perc of the identified avatars has unrealistic movement speeds (two to three
orders of magnitude higher speed than the average speed). We assume that these are
mainly due to the usage of long distance transportation systems such as the teleport.
We, thus, ignore such extremely high movements from the result. From this evaluation,
it is calculated that average speeds of avatars are 4.25~\wms (\ispa) and 6.39~\wms
(\ispb). \fref{fig:distance} illustrates that \lions show more itinerant
behavior than \tigers. This is likely due to the fact that \tigers spend more time
for communicating with other avatars in the virtual world, whereas \lions focus more
on hunting in the battle field. This is a reasonable inference because, for \lions,
a communication with other users does not interfere in the movement in the virtual
world as their in-game friends are within the talking distance.


\section{Conclusions}
\label{sec:conclusion}
In this paper, we have presented a thorough analysis of \wow game traffic
based on two sets of anonymized packet-level traces collected from a
European tier-1 ISP in two different time periods. This study is intended
to provide ISPs and academia for a better understanding of MMORPG user behavior
and characteristics of network traffic generated by such games. For this study,
we developed the protocol analyzer designed to identify \wow traffic from the
overall Internet traffic and to extract unencrypted messages,~\ie{logon messages
and movement messages}, from the classified \wow traffic.

From this analysis, we uncover general trends and characteristics of \wow
traffic. More precisely, we find that the protocol is modified during the two
years (2008 to 2010) in such a way that servers send less information or more
compressed information to clients through the logon connections, and servers
and clients send more information or more detailed information to each other
through game connections. We also report that the bandwidth utilization of 
such games is comparatively low and more than 60\perc of the total packets that clients
send to the server are for updating the avatar's coordinates.
Then, we examine differences between solitary users (\tigers) and
users who play the game together with other users behind the same middle box
(\lions). We find that \tigers tend to play the game longer than \lions, while
\lions travel longer distance in the virtual world than \tigers.

The major restriction we have encountered during this study is highly
encrypted part of the payload. We are of the belief that we can find
similarities among the same type of users if we examine social interactions,
\eg{chatting and trading}, between users. 
However, we do not intend to violate the privacy of \wow subscribers. Thus,
our future work includes the development of the message classification method
that identifies messages without the payload inspection. Furthermore, we plan
to carry out a survey in the user community for finding out reasons of such
differences between the two groups.

\bibliographystyle{IEEEtran}
\bibliography{wow}
\end{document}